\newcommand{\be}{\begin{equation}}
\newcommand{\ee}{\end{equation}}
\newcommand{\bea}{\begin{eqnarray}}
\newcommand{\eea}{\end{eqnarray}}
\newcommand{\beas}{\begin{eqnarray*}}
\newcommand{\eeas}{\end{eqnarray*}}
\newcommand{\bd}{\begin{displaymath}}
\newcommand{\ed}{\end{displaymath}}
\newcommand{\titj}{{\mbox{\boldmath$\tau$}}_i
                     \cdot{\mbox{\boldmath$\tau$}}_j}
\newcommand{\sisj}{{\mbox{\boldmath$\sigma$}}_i
                     \cdot{\mbox{\boldmath$\sigma$}}_j}
\newcommand{\lilj}{{\mbox{\boldmath$\lambda$}}_i
                     \cdot{\mbox{\boldmath$\lambda$}}_j}
\newcommand{\sinblr}{{\mbox{\boldmath$\sigma$}}_i
                     \cdot{\mbox{\boldmath$\nabla$}}_r}
\newcommand{\sjnblr}{{\mbox{\boldmath$\sigma$}}_j
                     \cdot{\mbox{\boldmath$\nabla$}}_r}
\def\d{ d$^\prime$ }

\documentstyle [11pt] {article}

\begin{document}

\title{ Pionic Decay of a Possible \d-Dibaryon }

\vskip 5mm
\author{ K. Itonaga,$^{\rm a,b}$  A. J. Buchmann,$^{\rm b}$  
Georg Wagner$^{\rm b}$ and Amand Faessler$^{\rm b}$ \\
 $^{\rm a}$ Laboratory of Physics, Miyazaki Medical College \\ 
 Kiyotake, Miyazaki 889-16, Japan \\
 $^{\rm b}$ Institute for Theoretical Physics, University of T\"ubingen \\
 Auf der Morgenstelle 14, D-72076 T\"ubingen, Germany }
\maketitle

\vskip 1cm

\begin{abstract}
	The pionic decay of a possible \d-dibaryon in the process 
\d $\to $ N + N + $\pi $ is studied in the microscopic quark shell 
model and with a single-quark transition operator describing the transition
 q $\to$ q + $\pi $. 
For the \d with quantum numbers 
$J^P = 0^-$, $ T = 0 $, we employ a six-quark shell-model wave function 
with a spatial $s^5p [51]_X$-configuration and with $N = 1$ 
harmonic oscillator quanta.
It is shown that the pionic decay width depends 
strongly on the mass and size of the \d. In the case that the 
calculated \d mass is close to the experimental one 
a small pionic decay  width of $\Gamma_\pi \approx 0.04$ MeV is obtained. 
This is an order of magnitude smaller than the experimental 
$\Gamma_{\pi}^{\rm exp} \simeq 0.5$ MeV. Two possibilities to improve 
the calculated width are suggested. The effect of the nonstatic 
correction term in the transition operator and the influence of the form 
factor at the decay vertex on the decay width are also discussed. 

\end{abstract}

\vskip 2cm

\noindent
1. INTRODUCTION

\vskip 2mm

	Dibaryons have a long history of more than three decades. The 
experimental search for dibaryons in the nucleon-nucleon (NN) channel has 
however not been successful yet.  On the other hand, more exotic 
baryon-number 
B = 2 systems, that is dibaryons 
which have a genuine six-quark $( n(q) -n(\bar q) = 6 )$ structure 
that cannot be described by hadronic degrees of freedom
are predicted by quantum chromodynamics (QCD).
If such an exotic dibaryon is proven to exist in even 
a single case, the consequences would 
be far reaching for understanding QCD 
in the low-energy domain of hadron and nuclear 
physics \cite{Seth85}. 

	Recently, new pionic double charge exchange (DCX) $(\pi^+, \pi^-)$ 
data have revived the interest in dibaryon physics.  
DCX measurements on a number of nuclear targets ranging from $^4{\rm He}$ to 
$^{56}{\rm Fe}$ show a sharp peak in the excitation 
function at an incident pion energy $T_\pi \simeq 50$ MeV 
and at forward angles $\theta_{\rm L} \approx 5^\circ$ \cite{Bil92,Clem94}. 
The location of the peak at $T_{\pi} \simeq 50$ MeV is nearly independent 
of the nuclear target. In the $^4{\rm He}$ - target case,  
the peak appears around $T_\pi \simeq 80 - 90$ MeV \cite{Clem94}
due to the additional 28 MeV of binding energy required to break up the 
$^4{\rm He}$ nucleus into an unbound four-neutron final state.

	The DCX process necessarily involves two nucleons due to charge 
conservation and is sensitive to short-range $NN$ correlations. 
In addition, because the peak position and the width 
are largely independent of the target, the observed resonance-like peak 
is likely to be connected with some elementary process. 
The invariant mass spectrum of the $p + p + \pi^-$ subsystem in the 
$ p+p \to p+p+ \pi^- + \pi^+ $ reaction also shows a peak at the 
same position as the $ (\pi^+, \pi^-)$ DCX experiments, which supports 
the interpretation of the observed peak as an elementary process 
\cite{Voro94}. 

	The resonance peak, which has been called \d-dibaryon, has an 
energy $M_{d'} = 2065$ MeV, spin-parity $J^P = 0^-$ and isospin 
$T = 0$ \cite{Bil92}.  The width of the peak is reported to be 
$ \Gamma_{medium} \approx 5$ MeV when the effect of the Fermi motion is 
subtracted \cite{Bil92}.  Such quantum numbers and such a small width are  
incompatible with the picture of an ordinary baryon-baryon ( NN$^*$, 
N$^*$N$^*$, etc. ) bound state. 

	Theoretically Mulders et al. predicted 
that six-quark states with quantum numbers
$J^P = 0^-, T = 0 $ should have a tetraquark ($q^4$) $-$
diquark ($q^2$) stretched structure and a mass $M \simeq 2100$ MeV 
within the MIT bag model \cite{Muld80}. 
Recently, the T\"ubingen group has studied the 
microscopic structure of a possible \d-dibaryon in the 
nonrelativistic quark model (NRQM). Two different bases 
have been employed,
the quark cluster model and the 
quark shell model \cite{Buch95,Wag95} in both cases paying 
due attention to the Pauli principle. 
In ref. \cite{Buch95} it has been shown that 
the nonrelativistic quark cluster model with a $q^4 - q^2$ 
-cluster configuration can account for the \d-mass of $M_{d^\prime} 
\simeq 2100$ MeV if the size parameter of the \d wave function 
is properly chosen \cite{Buch95}. In the translationally 
invariant quark shell model (TISM) \cite{Buch95,Wag95} the six-quark 
states with $J^P = 0^-, T = 0$ are evaluated in a model space 
including $N = 1$ and $N = 3$ 
harmonic oscillator (h.o.) excitations. The same effective 
interaction potentials as in the cluster model approach are used. 
The quark shell model gives similar \d-masses as the quark cluster model. 

	Another interesting problem is whether the NRQM 
can explain the observed decay width of the \d-resonance. This is 
the objective of this article. 
It is reported that the major part of 
the observed width $\Gamma_{medium}$ in DCX is due to the spreading width.
The spreading width is a result of the interaction of the \d with the 
nucleons in the nucleus; it is absent if the \d is formed in
an elementary reaction such as  $ p+p \to p+p+ \pi^- + \pi^+ $.
Since the \d is not allowed to decay into the $NN$ channel 
because of its quantum numbers $J^P = 0^-, T = 0$, the dominant decay 
mode of the \d would be the NN$\pi$ channel. 
The pionic decay width of the \d is estimated as small as 
$\Gamma_\pi \simeq 0.5$ MeV \cite{Bil92}. We evaluate the pionic decay 
width by employing the microscopic quark shell-model wave function of the 
\d as obtained in Ref. \cite{Wag95}. 

	In contrast to the isospin-0 assumption for the \d used here,
Valcarce et al. \cite{Val95} have discussed the possibility 
that the $J^P = 0^-$ resonance in the NN$\pi$ system 
seen in DCX experiments is an isospin-2 resonance in the nucleon-$\Delta$ 
channel.  Using a nonrelativistic quark model they have shown that the 
nucleon-$\Delta$ interaction is attractive in the  $J^P = 0^-$ channel
and that the decay width is explained if the calculated mass of the resonance 
approaches the experimental one.  

	This paper is organized as follows. 
In sect. 2 the basic expression for the pionic decay width is derived 
and its approximations are discussed. 
A refined treatment of the transition operator is also given in sect. 2. The 
quark shell model wave functions and the calculated \d-dibaryon mass 
are given in sect. 3. In sect. 4 the calculated results of the pionic decay 
widths are shown and the relation between the width and the \d-mass are 
discussed for several parameter sets. A summary is given in sect. 5.

\vskip 10mm

\noindent 
2. PIONIC DECAY WIDTH OF THE \d-DIBARYON

\vskip 3mm

\noindent 
{\it 2.1 Basic formula} 
\vskip 2mm
	The dominant decay mode of the \d-dibaryon with quantum 
numbers $J^P = 0^-, T = 0$ is the pionic decay, such as 
\d $\to$ p + p + $\pi^-$, p + n + $\pi^0$ and n + n + $\pi^+$. We treat 
the decay operator as well as the dibaryon and final nucleon states in 
terms of quark degrees of freedom. The decay of the \d-dibaryon arises 
from the single-quark transition $q \to q + \pi $, 
which is depicted in Fig. 1. 
 
\centerline {------------------}
\centerline { Fig. 1 }
\centerline {------------------}
 
\bigskip
\noindent
Fig. 1 \  The single-quark q $\to$ q + $\pi$ transition process.
${\mbox{\boldmath $p$}} \ ({\mbox{\boldmath $p$}}^\prime) $ denotes the 
initial (final) quark momentum and ${\mbox{\boldmath $k$}}$ is the pion 
momentum in the final state.  

\bigskip
\bigskip
The transition operator is expressed after nonrelativistic reduction as 

\be 
   {\cal O} = \sum_{j=1}^6 \frac{f_{\pi q}}{m_\pi}
({\mbox{\boldmath $\sigma$}}_j \cdot {\mbox{\boldmath $k$}})
({\mbox{\boldmath $\tau$}}_j \cdot {\mbox{\boldmath $\phi$}})
e^{-i{\mbox{\boldmath $k$}}\cdot ({\mbox{\boldmath $r$}}_j - 
{\mbox{\boldmath $R$}}_{\rm CM})}
\frac 1 {\sqrt{2E_\pi(2\pi)^3}} \ .
\ee

\noindent
Here, $f_{\pi q}$ represents the coupling constant at the qq$\pi$-vertex, 
$E_\pi (m_\pi)$ the pion energy (pion mass) and ${\mbox{\boldmath $\phi$}}$ 
the isovector of the pion field. The ${\mbox{\boldmath $\sigma$}} 
({\mbox{\boldmath $\tau$}})$ represent the spin (isospin) operator of
a single quark. 

	The nucleons and the pion emitted in the \d decay 
\d $\to$ N + N + $\pi$ have momenta 
$ {\mbox{\boldmath $ q$}}_1, {\mbox{\boldmath $ q$}}_2 $ and 
$ {\mbox{\boldmath $k$}}_\pi$, respectively, in the C.M. frame. 
We define 

\be
  {\mbox{\boldmath $Q$}} = {\mbox{\boldmath $q$}}_1 + 
{\mbox{\boldmath $q$}}_2 \ , \quad 
  {\mbox{\boldmath $q$}} = \frac {{\mbox{\boldmath $q$}}_1 
  - {\mbox{\boldmath $q$}}_2 } {2}  
\ee

\noindent
where ${\mbox{\boldmath $Q$}}$(${\mbox{\boldmath $q$}}$) is the total 
(relative) momentum of the two nucleons. Momentum and energy conservation 
of the decay process are expressed as 

\be
   0 = {\mbox{\boldmath $q$}}_1 + {\mbox{\boldmath $q$}}_2 + 
       {\mbox{\boldmath $k$}}_\pi = {\mbox{\boldmath $Q$}} + 
       {\mbox{\boldmath $k$}}_\pi  \ ,  
\ee

\be
   M_{d'} = 2M_N + \frac {{\mbox{\boldmath $Q$}}^2} {2(2M_N)} 
   + \frac {{\mbox{\boldmath $q$}}^2} {M_N} 
   + \sqrt{ m_\pi^2 + {\mbox{\boldmath $k$}}_\pi^2 }  \ , 
\ee 
\noindent
respectively, where $M_{d^\prime}$ represents the \d-dibaryon mass.

	The pionic decay width is expressed as 

\bea
  \Gamma_\pi^{\rm tot} & = & 2\pi \int d{\mbox{\boldmath $Q$}}
  d{\mbox{\boldmath $q$}}d{\mbox{\boldmath $k$}}_\pi 
  \delta \left( M_{d'} - 2M_N - \frac {{\mbox{\boldmath $Q$}}^2} {2(2M_N)} 
   - \frac {{\mbox{\boldmath $q$}}^2} {M_N} 
   - \sqrt{ m_\pi^2 + {\mbox{\boldmath $k$}}_\pi^2 } \right) 
   \nonumber \\
 &  & \times \ \  \delta( {\mbox{\boldmath $Q$}} + 
   {\mbox{\boldmath $k$}}_\pi ) 
  \frac 1 {2J_i + 1} \sum_{M_f} \sum_{M_i} \sum_{\tau_f} \nonumber \\
 &  & \times \ \  \vert < "N N" J_f M_f T_f \tau_f ; 
  {\mbox{\boldmath $\phi$}}_{\pi} \vert {\cal O} 
  \vert \Psi_{d'}, J_i M_i T_i \tau_i > \vert^2 \nonumber \\
 & = & 2\pi \int d{\mbox{\boldmath $q$}}d{\mbox{\boldmath $k$}}_\pi 
   \delta \left( M_{d'} - 2M_N  
   - \frac {{\mbox{\boldmath $k$}}_\pi^2} {4M_N} 
   - \frac {{\mbox{\boldmath $q$}}^2} {M_N} 
   - \sqrt{ m_\pi^2 + {\mbox{\boldmath $k$}}_\pi^2 } \right)  \nonumber \\
 &  & \times \ \  \frac 1 {2J_i + 1} \sum_{M_f} \sum_{M_i} \sum_{\tau_f} 
  \nonumber \\
 &  & \times \ \  \vert < "N N" J_f M_f T_f \tau_f ; 
  {\mbox{\boldmath $\phi$}}_{\pi} \vert {\cal O} 
  \vert \Psi_{d'}, J_i M_i T_i \tau_i > \vert^2  \nonumber \\
 &  & 
\eea

\noindent 
where the initial \d quantum numbers are referred to as $ J_i M_i T_i 
\tau_i $ and "$N N $" refers to the final two-nucleon state with quantum 
numbers $ J_f M_f T_f \tau_f $. To be specific we assume 
$J_i^P = 0^-, T_i = 0$ and $J_f^P = 0^+, T_f = 1$ due to the isovector 
nature $({\mbox{\boldmath $\tau$}})$ of the transition operator $\cal O$. 
Then the total pionic decay width can be written as 

\be
  \Gamma_\pi = \Gamma_{\pi^-} + \Gamma_{\pi^0} + \Gamma_{\pi^+} 
\ee

\noindent
and $\Gamma_{\pi^-}$ is expressed as 

\bea
  \Gamma_{\pi^-} 
 & = &  2\pi \int d{\mbox{\boldmath $q$}}d{\mbox{\boldmath $k$}}_\pi 
   \delta \left( M_{d'} - 2M_N  
   - \frac {{\mbox{\boldmath $k$}}_\pi^2} {4M_N} 
   - \frac {{\mbox{\boldmath $q$}}^2} {M_N} 
   - \sqrt{ m_\pi^2 + {\mbox{\boldmath $k$}}_\pi^2 } \right)  \nonumber \\
 & \times &  \vert < "N N" J_f=0^+, T_f=1, \tau_f=1 ; 
   \phi_{\pi^-} \ \vert {\cal O} \ \vert 
   \nonumber \\
 &  & \qquad \qquad \qquad \qquad \qquad \qquad 
   \Psi_{d'}, J_i=0^-, T_i=0, \tau_i=0 > \vert^2  \ . 
\eea

\noindent
Similar expressions hold for $\Gamma_{\pi^0}$ and $\Gamma_{\pi^+}$. 
Owing to isospin invariance, one has

\be
  \Gamma_{\pi^{-}} \equiv \Gamma_{\pi^0} \equiv \Gamma_{\pi^+} 
\ee

\vskip 5mm
\noindent 
{\it 2.2 Restriction to dominant intermediate six-quark states} 
\vskip 2mm

	The decay transition matrix element can be evaluated by inserting 
a complete set of  six-quark states as 

\bea 
< &"N N"& J_f=0^+, T_f=1, \tau_f=1 ; 
   \phi_{\pi^-} \vert {\cal O} \vert 
   \Psi_{d^\prime}, J_i=0^-, T_i=0, \tau_i=0 >  \nonumber \\
 & = & \sum_n  < "N N" J_f=0^+, T_f=1, \tau_f=1  \vert 
  \psi_{6q,n} > \nonumber \\
 &  & \qquad \times  < \psi_{6q,n}, \phi_{\pi^-} 
  \vert   {\cal O} \vert \Psi_{d^\prime}, J_i=0^-, T_i=0, \tau_i=0 >  \ . 
\eea 

	The first factor on the r.h.s. of Eq. (9) is the overlap between 
the antisymmetrized two-nucleon state and the intermediate six-quark 
state of the n-th excitation. The second factor which contains the decay 
dynamics connects the \d-dibaryon and the intermediate six-quark 
state accompanying the emitted $\pi^-$-meson wave. The quark shell model 
$ s^5p [51]_X $-configuration with $N = 1$ h.o. excitation 
is used for $\Psi_{d^\prime}(J = 0^-,T = 0)$.

	The following arguments restrict the sum over intermediate 
states in Eq. (9).
 First, the pionic decay proceeds through a one-body operator. 
This means that only such six-quark configurations in which 
a single orbit ($s$ or $p$) is changed from the initial $s^5p$-configuration 
are allowed. 
Second, those intermediate six-quark states 
which have a large overlap with asymptotic $NN$ components
(large fractional parentage factors)
will give an appreciable contribution to the transition matrix element. 
In view of this, the two lowest-lying six-quark states are 
adopted as intermediate states in this work. 
These are the $0 \, \hbar\omega$ 
$\psi_{6q}(s^6[6]_X, J = 0^+, T = 1)$ six-quark state 
with $N = 0$ and  the $2 \, \hbar\omega$ excited state 
 $\psi_{6q}(s^4p^2[42]_X, J = 0^+, T = 1)$  with $N = 2$. 
A more detailed discussion of these wave functions 
as well as of the \d wave function is given in sect. 3.

\vskip 5mm
\noindent 
{\it 2.3 Nonstatic correction and form factor in the transition operator} 
\vskip 2mm
	One can go beyond the static limit of Eq. (1) 
and consider the next-to-leading order relativistic correction in the
transition operator $ \cal O $.
The nonstatic correction term at the qq$\pi$-vertex 
in Fig. 1 might be important in the present quark model description of 
the \d decay because the  
ratio of pion and quark masses is not small. The nonstatic correction 
of order $ {\cal O}(\frac {m_\pi} {m_q})$ can be taken into account 
by modifying the momentum-dependent part of the operator $ \cal O$ of 
Eq. (1) as follows

\be
  ( {\mbox{\boldmath $\sigma$}}_j \cdot {\mbox{\boldmath $k$}} )
  \to  
  \left( {\mbox{\boldmath $\sigma$}}_j \cdot ({\mbox{\boldmath $k$}} - 
  \frac {m_\pi} {2m_q} \frac { {\mbox{\boldmath $p$}}^\prime + 
  {\mbox{\boldmath $p$}} } {2}) \right) 
\ee

\noindent
where ${\mbox{\boldmath $p$}} \ ({\mbox{\boldmath $p$}}^\prime) $ is the 
initial (final) quark mo\-men\-tum \cite{Eric88}. The quark 
mo\-men\-tum-de\-pen\-dent term in Eq. (10) is often referred 
to as the recoil correction. 

	Another refinement is due the finite size of the 
qq$\pi$-vertex in Fig. 1. Usually, this is treated by defining a 
renormalized momentum-dependent coupling in place 
of the $f_{\pi q}$ in Eq. (1)  

\be
      f_{\pi q}  \to  f_{\pi q} \cdot \left[ \frac {\Lambda^2} 
           {\Lambda^2 + {\mbox{\boldmath $k$}}^2 } \right]^{1/2}, 
\ee
\noindent
consistently with the pion-exchange potential \cite{Fer86}. The cut-off mass 
$\Lambda$  measures the extension of the core-size of the 
quark-pion interaction.

\vskip 10mm
\noindent
3. THE \d-DIBARYON MASS AND QUARK SHELL MODEL WAVE FUNCTIONS 

\vskip 3mm

\noindent 
{\it 3.1 The \d-dibaryon mass and wave function} 
\vskip 2mm

	The \d-dibaryon mass and wave functions which have been studied in 
Refs. \cite{Buch95,Wag95} are employed in the present calculation 
of the pionic decay rate of the \d. The basic ingredients and the results of 
Refs. \cite{Buch95,Wag95} which are essential for the present calculation 
are thus briefly recapitulated.

	Although quark dynamics is known to be described 
by QCD, the many-quark system ($n(q) \geq 3$)  
has not been solved successfully from first principles 
except for calculations in the lattice gauge theory. In the low-energy 
domain, the nonrelativistic constituent quark model works well in 
accounting for the regularities of single hadron spectra. 

	The translationally invariant quark shell model Hamiltonian which 
includes the effective quark-quark interaction is written as 

\be
 H = \sum_{i=1}^n (m_q + \frac { {\mbox{\boldmath $p$}}_i^2} {2m_q} ) 
   - \frac { {\mbox{\boldmath $P$}}^2} {n(2m_q)} 
   + \sum_{i<j}^n V^{Conf}({\mbox{\boldmath $r$}}_i,{\mbox{\boldmath $r$}}_j) 
   + \sum_{i<j}^n V^{res}({\mbox{\boldmath $r$}}_i,{\mbox{\boldmath $r$}}_j) .
\ee 

\noindent
The confinement potential $V^{Conf}$ is necessary; it is assumed to be linear 
\cite{Wag95} or quadratic \cite{Buch95} in 
$\vert {\mbox{\boldmath $r$}} \vert = 
\vert {\mbox{\boldmath $r$}}_i - {\mbox{\boldmath $r$}}_j \vert $, i.e., 

\bea
 V^{Conf}({\mbox{\boldmath $r$}}_i,{\mbox{\boldmath $r$}}_j) 
 &=&   - \frac {1} {4} \lilj 
( a_c \vert {\mbox{\boldmath $r$}}_i - {\mbox{\boldmath $r$}}_j \vert - C ) ,
 \\
 & {\rm or} &    \nonumber \\
 &=&   - a_c^\prime  {\mbox{\boldmath $\lambda$}}_i \cdot 
    {\mbox{\boldmath $\lambda$}}_j  
 ({\mbox{\boldmath $r$}}_i - {\mbox{\boldmath $r$}}_j)^2,  
\eea 

\noindent
where ${\mbox{\boldmath $\lambda$}}_i$ is the SU(3)$_{color}$ matrix of the 
$i$-th quark. The residual interaction $V^{res}$ is composed of the 
chiral field ( ${\mbox{\boldmath $\pi$}}$ and ${\mbox{\boldmath $\sigma$}}$ ) 
potential and the one-gluon exchange potential.  
The former one-pion and one-sigma exchange potentials are closely related 
to the spontaneous breaking of chiral symmetry of QCD \cite{Kus91,Fer93}. 
The latter one-gluon exchange potential is responsible for the short-range 
part of the interaction \cite{DeR75}. 
They are expressed as 

\be
 V^{OPEP}({\mbox{\boldmath $r$}}_i,{\mbox{\boldmath $r$}}_j) 
  = \frac {g_{\pi q}^2} {4\pi} \frac 1 {4m_q^2} \frac {\Lambda^2} 
 {\Lambda^2 - m_\pi^2} (\titj)(\sinblr)(\sjnblr) 
 \left( \frac {e^{-m_\pi r}} {r} - \frac {e^{-\Lambda r}} {r} \right) \ ,
\ee 

\be
 V^{OSEP}({\mbox{\boldmath $r$}}_i,{\mbox{\boldmath $r$}}_j) 
  = - \frac {g_{\sigma q}^2} {4\pi} \frac {\Lambda^2} 
 {\Lambda^2 - m_\sigma^2} 
 \left( \frac {e^{-m_\sigma r}} {r} - \frac {e^{-\Lambda r}} {r} \right) 
\ee 

\noindent
with

\be
 \frac {g_{\pi q}^2} {4\pi} = \frac {g_{\sigma q}^2} {4\pi} , \quad 
 m_\sigma^2 \simeq (2m_q)^2 + m_\pi^2 , \quad 
 \Lambda_\pi = \Lambda_\sigma = \Lambda , 
\ee

\noindent
and

\be
 V^{OGEP}({\mbox{\boldmath $r$}}_i,{\mbox{\boldmath $r$}}_j) 
  =   \frac {1} {4} \alpha_s \lilj 
 \left\{ \frac 1 r - \frac {\pi} {m_q^2} ( 1 + \frac 2 3 \sisj ) 
 \delta ({\mbox{\boldmath $r$}}) \right\} , 
\ee

\noindent
where ${\mbox{\boldmath $\sigma$}}_i $ is a spin operator. The vertex form 
factor $ ( \frac {\Lambda^2} { \Lambda^2 + {\mbox{\boldmath $k$}}^2 } )^{1/2}$ 
is used in deriving Eqs. (15) and (16). The relations 
$ g_{\pi q}^2 = f_{\pi q}^2 ( \frac {2m_q} {m_\pi} )^2 $ and 
$ f_{\pi q} = \frac 3 5 f_{\pi N} $ hold, where $f_{\pi N}$ is the NN$\pi$ 
coupling constant. 
The parameter sets used in this work are listed in Table I for
both linear and quadratic confinement potentials.
For sets 1-2 the parameters are determined by 
fitting the light baryon masses ( N, $\Delta$, N$^*$(1535) ), and by 
satisfying the nucleon stability condition $\partial M_N /\partial b_N=0$
and the approximate constraints for $\Lambda$ and $m_q$. 
Parameter sets 3-5 fit the $N$ and $\Delta$ masses but underestimate the
N$^*$(1535) mass by about 300 MeV. The latter 3 parameter sets all have 
a small confinement strength $a_c$.

\vskip 1cm
\noindent
Table I  \ Quark shell model parameters 
and the calculated \d-dibaryon mass $M_{d^\prime}$. The $b_N$ and 
$b_6$ are the h.o. size parameters of the final nucleon and the \d-dibaryon 
wave functions, respectively. $m_q$ is the constituent quark mass. $\alpha_s, 
a_c, C, a_c^\prime$ and $\Lambda$ are the potential parameters.
Set 1-3: The values of the parameters and the corresponding 
\d-dibaryon masses for a linear confinement potential (Eq. (13)).
Set 4-5: The values of the parameters and the corresponding \d-dibaryon masses 
for a quadratic confinement potential (Eq. (14)).

\vskip 2mm
\begin{tabular}{*{10}{c}} 
\hline 
\hline
\thinspace
    & $b_N$ & $\alpha_s$ & $a_c$  & $C$ & $m_q$  & $\Lambda$  &  & 
$b_6$ & $M_{d^\prime}$ \\
SET & [fm] &  & [${\rm MeV} \cdot {\rm fm}^{-1}$] & [MeV] & 
[MeV] & $[{\rm fm}^{-1}]$ &  & [fm] & [MeV]  \\ 
\hline
\thinspace 
 1 & 0.45 & 0.127 & 462 & 549 & 338 & 5.07 &  & 0.59 & 2705  \\
 2 & 0.47 & 0.074 & 423 & 500 & 296 & 7.60 &  & 0.65 & 2680  \\
 3 & 0.60 & 0.816 &  25 & $-32$ & 313 & 10.14 &  & 1.24 & 2162  \\
\hline
\hline
\thinspace
    & $b_N$ & $\alpha_s$ & $a_c^\prime$  &  & $m_q$  & $\Lambda$  &  & 
$b_6$ & $M_{d^\prime}$ \\
SET & [fm] &  & [${\rm MeV} \cdot {\rm fm}^{-2}$] &  & 
[MeV] & $[{\rm fm}^{-1}]$ &  & [fm] & [MeV]  \\ 
\hline
\thinspace 
 4 & 0.595 & 0.958 & 13.97 & $-$ & 313 & 4.20 &  & 0.78 & 2484  \\
 5 & 0.595 & 0.958 & 5.0 & $-$ & 313 & 4.20 &  & 0.95 & 2112  \\
\hline
\end{tabular}

\vskip 1cm
	The lowest-lying harmonic oscillator state 
which is compatible with \d  quantum numbers $J^P = 0^-$ and $T = 0$ 
involves one quantum excitation $N=1$ and is uniquely 
expressed as 

\bea
 \Psi_{d^\prime}( \ N=1, s^5p[51]_X, (\lambda \mu)=(10), & L=1,& S=1, J=0^-, 
 T=0, \nonumber \\
 & & [2211]_{CT}, [222]_C \ ). 
\eea

\noindent
Here,  $[...]_X, [...]_{CT}$ and $[...]_C$ specify the space-, color-isospin- 
and color- symmetry, respectively, and $(\lambda \mu)$ represents the 
Elliott SU(3) label of h.o. quanta. The h.o. size parameter of the 
$\Psi_{d^\prime}$ wave function is $b_6$. Other higher excited configurations 
with $N = 3$ excitations and $[42]_X$ symmetry are considered in 
Ref.\cite{Wag95}. A lower 
\d-mass is obtained in the configuration mixing calculation but the mixing 
amplitudes are small. Therefore, in the present paper, we adopt only the 
lowest configuration of Eq. (19) for the \d-wave function. The size 
parameter $b_6$ is determined by minimizing the \d-mass.  Calculated 
\d-masses $M_{d^\prime}$ and corresponding size parameters $b_6$ are shown in 
Table I. It is noticed that in the cases of weaker 
confinement potentials (set 3 \& 5), the size parameters $b_6$ become large 
and consequently lower dibaryon masses $M_{d^\prime}$, which are 
comparable with the experimental resonance energy of 2065 MeV, are obtained.

\vskip 5mm
\noindent 
{\it 3.2 The intermediate six-quark wave functions and the final 
two-nucleon state} 
\vskip 2mm
 
	Among all possible intermediate states only the 
lowest two six-quark states with $J^P = 0^+, T = 1$ are adopted in Eq. (9) 
for the evaluation of the pionic decay width of the \d. The criteria for 
selecting these six-quark states have already been mentioned in sect. 2.2. 
The lowest two states with $N = 0$ and $N = 2$ h.o. excitations are 
expressed as 
\bea 
 \psi_{6q,1} = \psi_{6q} ( \ N = 0, s^6[6]_X, (\lambda \mu) = (00), 
 & L=0,& S=0, J=0^+, T=1,  \nonumber \\
 & &  [222]_{CT}, [222]_C \ ) 
\eea

\noindent
and

\bea
 \psi_{6q,2} = \psi_{6q} ( \ N = 2, s^4p^2[42]_X, (\lambda \mu) = (20), 
 & L=0,& S=0, J=0^+, T=1,  \nonumber \\
 & &  [42]_{CT}, [222]_C \ ) \ . 
\eea
\noindent
The probability amplitudes for finding the two-nucleon component
in these two six-quark states (fractional parentage coefficients)
are $ \sqrt {1/9}$ and 
$ - \sqrt {1/25}$, respectively \cite{Har81,Gloz93}, if a common 
h.o. size parameter is assumed for $\psi_{6q,i} (i=1,2)$ 
and if the nucleon has an $s^3$-configuration.

	The final outgoing two-nucleon state is approximately written as 

\bea
  \vert \ "NN" ; &J& = 0^+, T=1 >  \nonumber \\
  &=& \frac 1 { \sqrt {2}} \vert \ \psi_N \psi_N \frac 1 { (2\pi)^{3/2}} 
 [ e^{i{\mbox{\boldmath $q$}}{\mbox{\boldmath $r$}} } 
  - (-1)^{S+T} e^{-i {\mbox{\boldmath $q$}}{\mbox{\boldmath $r$}} } ] 
  \chi_{M_S}^{S=0} \xi_{M_T}^{T=1} \ ; \nonumber \\
  & &  \qquad \qquad \qquad L=0, S=0, J=0^+, T=1 \ > \ . 
\eea

\noindent
In Eq. (22), an exchange of three-quarks between nucleons is taken into 
account. $\psi_N $ represents a free nucleon and its size parameter 
is given by $b_N$ in Table I.

\vskip 10mm
\noindent
4. RESULTS AND DISCUSSION

\vskip 3mm

	The expression for the pionic decay width of the \d is listed in the
Appendix. The analytic form of this expression 
is an advantage of the present model; it facilitates the discussion and 
interpretation of our results. 
The decay widths $\Gamma_{\pi^-}$ and $\Gamma_\pi$ 
are evaluated by adopting the wave function $\Psi_{d^\prime}$ of Eq. (19) 
and several sets of parameters for the linear confinement 
(sets $1 - 3$) and the quadratic confinement (sets 4 \& 5) potential cases. 
The calculated decay widths are shown in Table II under the abbreviated 
symbols S, S+NSC and S+NSC+FF, where S refers to a calculation 
employing the static decay operator of Eq. (1), 
NSC includes the nonstatic correction term of Eq. (10), 
and FF refers to a calculation using the form factor of Eq. (11). 
In Table II, two numbers for $\Gamma_{\pi^-}$ are listed in each column 
S, S+NSC and S+NSC+FF in order to show the effect of the intermediate 
six-quark states in Eq. (9); the numbers without parenthesis are 
widths evaluated by adopting both configurations $\psi_{6q,1}$ and 
$\psi_{6q,2}$ (abbreviated as I \& II, respectively) and the numbers 
with parenthesis [ ] are the widths evaluated with configuration I 
only.

\vskip 1cm
\newcommand{\lw}[1]{\smash{\lower2.ex\hbox{#1}}}
\renewcommand{\arraystretch}{1.2}

\noindent
Table II \ 
Calculated $\pi^-$-decay width $\Gamma_{\pi^-}$ 
and the total $\pi$-decay width $\Gamma_\pi$ of the \d-dibaryon 
for the adopted parameters sets 1 - 5. The experimental data \cite{Bil92} is 
also shown. The symbol S denotes the calculation with use of the operator 
${\cal O}$ of Eq. (1) in the static approximation, S+NSC the calculation 
with the operator ${\cal O}$ with inclusion of the nonstatic correction 
term of Eq. (10) and S+NSC+FF the calculation with the operator ${\cal O}$ 
with inclusion of nonstatic correction term and the form factor of Eq. (11). 
The numbers without the parenthesis in each column of S, S+NSC and S+NSC+FF 
of $\Gamma_{\pi^-}$ are the calculations when the configurations I and II 
($\psi_{6q,1}$ and $\psi_{6q,2}$) are adopted for the intermediate 
six-quark states. The numbers with the parenthesis [ ] are the 
calculations when only the configuration I is adopted for the intermediate 
six-quark state. The right-most column lists the calculated decay width
$\Gamma_\pi = 3 \Gamma_{\pi^-}$.

\vskip 2mm
\begin{tabular}{*{8}{c}}
\hline
\hline
 \lw{ SET} & \lw{$b_6$[fm]} & \lw{$M_{d^\prime}$[MeV]} & 
  \multicolumn{3}{c}{$\Gamma_{\pi^-}$ [MeV]} &  & 
 {$\Gamma_{\pi}$} [MeV] \\
  \cline{4-6}  \cline{8-8} 
   &  &  & S & S+NSC & S+NSC+FF &  & S+NSC+FF \\
\hline
 1 & 0.59 & 2705 & 8.16 [8.83] & 7.16 [7.85] & 5.33 [5.82] & & 16.0 \\
 2 & 0.65 & 2680 & 6.70 [7.15] & 6.34 [6.84] & 5.53 [5.95] & & 16.6 \\
 3 & 1.24 & 2162 & 0.03 [0.03] & 0.01 [0.02] & 0.01 [0.02] & & 0.04 \\
 4 & 0.78 & 2484 & 4.22 [4.52] & 3.74 [4.06] & 2.90 [3.13] & & 8.69 \\
 5 & 0.95 & 2112 & 0.02 [0.02] & 0.01 [0.01] & 0.01 [0.01] & & 0.04 \\
EXP. &  &  &  &  &  &  & $\simeq$ 0.5 \\
\hline
\end{tabular}

\vskip 1cm
 	Let us first discuss the decay width calculated with the operator 
$\cal O$ in the static approximation in Eq. (1). The calculated numbers 
are quoted in column S in Table II. The decay amplitude which connects the 
dibaryon state $\Psi_{d^\prime}$ and the final two-nucleon and one-pion state 
is the essential 
ingredient in determining the pion decay width. 
The comparison of the calculated decay width with experiment 
provides a severe test of the validity of various model assumptions.

	The pion decay width is very sensitive to the 
\d-dibaryon mass $M_{d^\prime}$ which essentially determines the phase 
space of the three-body NN$\pi$-decay. The decay kinematics is given in 
Eqs. (3) and (4). Due to the larger phase space,
a large $M_{d^\prime}$ leads to a large 
pionic decay width $\Gamma_{\pi^-}$ and $\Gamma_\pi$.
This is seen in the cases of sets 1 \& 2 (linear 
confinement) and set 4 (quadratic confinement) in Table II. If the 
calculated $M_{d^\prime}$ is close to the experimental value 2065 MeV 
as in set 3 (linear confinement) and set 5 (quadratic confinement), 
a small $\Gamma_{\pi^-}$ ($\Gamma_\pi$) is obtained. 

	Another factor which affects the decay width is the overlap
factor between the final two-nucleon state in free space and the two 
three-quark-states ($s^3$) in the six-quark configuration. The 
overlap-squared is expressed as 

\be
\label{overlap}
    \left ( \frac { 2(b_6 / b_N)} { 1+ (b_6 / b_N)^2 } \right )^{12},
\ee

\noindent
where $b_6 (b_N) $ is the  h.o. size parameter of the six-quark (free-nucleon) 
state. The overlap-squared values for sets 1, 2 and 4 are almost similar and 
as large as $ \approx 0.6$, while for sets 3 \& 5 the values are as small 
as 0.05 and 0.28, respectively. The latter small values for sets 3 \& 5 
are due to the very different sizes of $b_6$ from $b_N$ as seen in Table I. 

	In view of the above two key-factors, large decay widths 
$\Gamma_{\pi^-}$ are obtained for sets 1, 2 and 4, while much smaller 
$\Gamma_{\pi^-}$ are obtained for sets 3 \& 5 (see the column S in Table II). 

	One can see the role of adding the configuration II from the 
comparison of the two numbers without and with parenthesis [ ]. 
The inclusion of configuration II reduces the decay 
width $\Gamma_{\pi^-}$ by about $7 - 10$ \% compared with the 
values for configuration I only. 
This reduction can be physically understood. It is due to the 
destructive interference between the $s^6$ and 
the $s^4p^2$ six-quark component of the $NN$ wave function at short distances
which results in a smaller intermediate six-quark amplitude \cite{Fae83}.
The effect of the excited $N=2$ 
intermediate six-quark state
$\psi_{6q,2}$ with an $s^4p^2[42]_X $-configuration 
turns out to be significant but does not change
the results qualitatively. This is because the transition amplitude which 
connects $\Psi_{d^\prime}$ with $\psi_{6q,2}$ and $\phi_\pi$ is small 
compared with that between $\Psi_{d^\prime}$ and $\psi_{6q,1}$ and 
$\phi_\pi$. 

	We proceed to discuss the effect of the nonstatic correction 
(NSC) term of the transition operator to the width. Since in the 
preceeding paragragh the contribution of $\psi_{6q,2}$ to the amplitude is 
found to be rather small, we evaluate the NSC effect only for the 
transition from $\Psi_{d^\prime}$ to $\psi_{6q,1}$ and $\phi_\pi$. 

	The right-hand-side operator in Eq. (10) is rewritten as 

\bea
  ( {\mbox{\boldmath $\sigma$}}_j \cdot {\mbox{\boldmath $k$}} )
 & - &  ( \frac {m_\pi} {2m_q} ) 
  \frac { {\mbox{\boldmath $\sigma$}}_j \cdot 
  ( {\mbox{\boldmath $p$}}^\prime + {\mbox{\boldmath $p$}} ) } {2}  
 \nonumber \\
 & = & ( {\mbox{\boldmath $\sigma$}}_j \cdot {\mbox{\boldmath $k$}} ) 
  + \left \{ \frac 1 2 ( \frac {m_\pi} {2m_q} ) 
  ( {\mbox{\boldmath $\sigma$}}_j \cdot {\mbox{\boldmath $k$}} ) 
  - ( \frac {m_\pi} {2m_q} ) 
  ( {\mbox{\boldmath $\sigma$}}_j \cdot {\mbox{\boldmath $p$}} ) \right \} \ . 
\eea

\noindent
The first term 
$ ( {\mbox{\boldmath $\sigma$}}_j \cdot {\mbox{\boldmath $k$}} ) $ is the 
operator in the static approximation and the second term the nonstatic 
correction. The transition amplitudes between $\Psi_{d^\prime}$ and 
$\psi_{6q,1}$ and $\phi_\pi$ are expressed symbolically in the following 
form : For the operator 
$({\mbox{\boldmath $\sigma$}} \cdot {\mbox{\boldmath $k$}})$ 
it reads 

\be
 < \psi_{6q,1}, \phi_\pi  \ \vert \ 
  ( {\mbox{\boldmath $\sigma$}} \cdot {\mbox{\boldmath $k$}} ) \cdot \cdot 
  \cdot \vert \ \Psi_{d^\prime} >  \sim {\rm const} \cdot (b_6 k^2) 
  e^{ -\frac 5 {24} (b_6 k)^2 } \ , 
\ee

\noindent
while for the operator due to the nonstatic correction it reads 

\bea
 < \psi_{6q,1}, \phi_\pi  & \vert & 
  \left \{ \frac 1 2 ( \frac {m_\pi} {2m_q} ) 
  ( {\mbox{\boldmath $\sigma$}} \cdot {\mbox{\boldmath $k$}} ) 
  - ( \frac {m_\pi} {2m_q} ) 
  ( {\mbox{\boldmath $\sigma$}} \cdot {\mbox{\boldmath $p$}} ) \right \} 
  \cdot \cdot \cdot \vert \ \Psi_{d^\prime} >  \nonumber \\
  & \sim & {\rm const} \cdot ( \frac {m_\pi} {2m_q} ) 
  ( \beta + \beta^\prime (b_6 k)^2 ) \frac 1 {b_6} 
  e^{ -\frac 5 {24} (b_6 k)^2  } \ , 
\eea

\noindent
where $b_6$ is the size parameter of $\Psi_{d^\prime}$ and $\psi_{6q,1}$. 

	Apart from the exponential factor, the following features are 
noticed in Eqs. (25) and (26). In the domain of large $k$, the term 
proportional to $k^2$ in Eq. (25) contributes most, while  in the 
small $k$ region the nonstatic correction dominates over 
the static approximation term due to the presence of the $k$-independent 
term in Eq. (26). 

	Combining Eqs. (25) with (26), the effect of incorporating the 
nonstatic correction term in the operator is written in a factorized 
form as [NSCI.]. An explicit expression for [NSCI.] is given in the 
Appendix. Note that $k$ is 
replaced by $k_0$ in the Appendix.  The factor [NSCI.] equals to 1 if no NSC 
is considered. The factor [NSCI.] changes appreciably as $k$ varies. 
The NSC reduces the decay width $\Gamma_{\pi^-}$ roughly 
by $ \sim 10$ \% for sets 1, 2 and 4 compared with the corresponding 
values in column S. However, for set 3 the NSC is quite important and the 
width $\Gamma_{\pi^-}$ is reduced to half the value in column S,
while for set 5 the NSC changes $\Gamma_{\pi^-}$ only slightly.    

 	Finally, the form factor (FF) effect is discussed. The FF at the 
qq$\pi$-vertex is of square-root type as shown in the Appendix. It 
is most effective for large momentum transfers $k_0$. Thus 
a strong FF induced suppression of the decay width is seen for sets 1, 2 
and 4 in which large momentum transfers $k_0$ are involved. On the 
contrary, the FF effect is negligible for sets 3 \& 5 
where the momentum transfer to the pion is small. 

	The final results for the total pionic decay width 
$\Gamma_\pi$ are listed in the right-most column
under the symbol S+NSC+FF, which are to be compared with the 
experimentally reported width $\Gamma_\pi^{\rm exp} \simeq 0.5$ MeV. The 
calculated $\Gamma_\pi$ for sets 1, 2 and 4 are 
too large in comparison with experiment. With $\Psi_{d^\prime}$ and 
the parameters for sets 1, 2 and 4, the calculated dibaryon masses 
$M_{d^\prime}$ are $600 - 400$ MeV above the resonance 
mass 2065 MeV. This overestimation of the \d mass is mostly 
responsible for the large pionic decay widths. 
The calculated $\Gamma_\pi$ for 
sets 3 \& 5 are both 0.04 MeV which is an order of magnitude too small 
compared with the experimental data. Although the dibaryon masses 
$M_{d^\prime}$ close to the experimental one are obtained 
by using $\Psi_{d^\prime}$ and the parameter sets 3 and/or 5, 
the corresponding sizes of the dibaryon $b_6$ for these cases are large 
compared with the size parameter of the nucleon $b_N$. 
Thus the overlap factor involved 
in the decay width calculation comes out too small for sets 3 \& 5 
( 0.05 \& 0.28, respectively ), which in turn results in a too small 
decay width.     
   
	We now come to a critical discussion of these results. 
In order to explain the pionic decay width of the 
\d-dibaryon, first of all the experimental \d-mass $M_{d^\prime}$ has 
to be satisfactorily reproduced by the wave function    
$\Psi_{d^\prime}$. With the parameters of sets 1, 2 and 4 
which produce large \d-masses far from the experimental value 
it seems impossible to explain the decay width $\Gamma_\pi^{\rm exp} $.  
The \d mass can be explained by  $\Psi_{d^\prime}$ with parameter 
sets 3 \& 5. It is however noted that very weak confinement potentials 
are adopted for these cases and the low \d-masses $M_{d^\prime}$ close to 
the experimental resonance energy are obtained at the price of having an 
extended radial \d wave function $\Psi_{d^\prime}$ with a large $b_6$. 
Thus, although the \d-mass $M_{d^\prime}$ is correctly reproduced 
for sets 3 \& 5, the size-parameter mismatch between the final 
free nucleons ($b_N$) and the three-quark-states ($s^3$) 
in $\Psi_{d^\prime}$ ($b_6$) results in a very small overlap 
and consequently leads to a small pionic decay width. 

	Suppose that a certain Hamiltonian
would produce the \d mass of 2112 MeV (set 5) 
with a size parameter $b_6 = 0.65$ fm (0.595 fm) 
in the wave function $\Psi_{d^\prime}$. In this case one would 
obtain $\Gamma_{\pi}$ as 0.36 MeV (0.40 MeV) which is close to 
the experimental $\Gamma_{\pi}^{\rm exp} \simeq 0.5$ MeV. 

	Two possibilities are suggested to achieve a small decay width. 
One possibility is to study the effect of 
additional quark-quark interactions which are not considered in the 
present work and which might explain the \d mass 
correctly. As we have argued above, the corresponding 
wave function $\Psi_{d^\prime}$ should not be much wider than the 
wave function
of a single free baryon in order that the overlap factor of Eq.(\ref{overlap})
remains close to unity. The second possibility is concerned with an improved 
treatment of the final state.  In the present work the relative motion 
of the final state nucleons and of the pion are simply assumed to be 
described by plane waves. In a more complete treatment, the final state
should be calculated with the same Hamiltonian that is used to calculate the 
mass of the \d; in other words an NN$\pi$ scattering wave function should
be used in the calculation of the decay amplitude.
If such final state wave functions are 
incorporated in the pionic decay process, for example in the cases of 
sets 3 \& 5, an enhanced decay width would be obtained. 
This enhancement is expected because the 
relative wave functions are distorted and contain high momentum components 
which will in turn lead to an enhanced decay width \cite{Ito88}. 
It is noted that Schepkin et al. \cite{Sch93} have estimated the 
enhancement factor of the pionic decay width ($\eta \sim 5$ in their 
model) when they considered the final state interaction for the decaying 
nucleons.

\vskip 10mm
\noindent
5. SUMMARY 
\vskip 3mm

	The pionic decay width of a possible \d-dibaryon found in 
$(\pi^+,\pi^-)$ DCX reactions on many nuclear targets is studied in 
the microscopic quark shell model. We employ a single-quark 
transition operator which expresses the conversion of a single-quark state 
into another single-quark one by emitting a pion. We adopt the shell 
model wave functions devised by the T\"ubingen group.  They have investigated 
the \d-mass and its structure in the translationally invariant quark shell 
model which includes the quark-quark interactions such as the confinement
potential, the chiral field ({\mbox{\boldmath$\pi$}}- 
and {\mbox{\boldmath$\sigma$}}-exchange) 
and the one-gluon-exchange potential.  In this work, the orbital 
\d-wave function with $J^P = 0^-$, $T=0$ is given 
by a $s^5p[51]_X$-configuration with  $N = 1$ 
harmonic oscillator excitation.
The pionic decay operator in its simplest form is obtained by the 
non-relativistic reduction of the  qq$\pi$-vertex and by imposing the 
static approximation on the operator.  A more refined transition operator 
which includes the non-static (recoil) correction term and also the form 
factor at the  qq$\pi$-vertex is discussed.

	The pionic decay transition matrix element is evaluated in
our model as the sum of products of two amplitudes; the first amplitude  
describes the transition from the \d-state to the intermediate 
six-quark state and a pion, the second amplitude represents the 
overlap between the intermediate six-quark state and the final two-nucleon 
state. The sum extends over important intermediate six-quark configurations.
The calculated decay width turns out to be strongly dependent 
on the two basic properties of the \d, i.e. its mass $M_{d^\prime}$ and its 
size $b_6$.  The \d-dibaryon mass essentially determines the phase space of 
the pionic decay and hence controls the decay width.  
The size of \d strongly affects the overlap between the 
intermediate six-quark state and the final two-nucleon state. 
The pionic decay width also depends on the number of intermediate states 
included and on the treatment of the decay operator.

	The pionic decay widths and masses of the \d 
are evaluated for microscopic \d wave functions and for several sets of 
model parameters \cite{Buch95,Wag95}.
In the case that the calculated \d mass is $600 - 400$ MeV
above the experimental value 2065 MeV, the decay width 
$\Gamma_{\pi} \approx 16-8$ MeV is much too large
compared to the experimental result. On the other hand, a decay width 
of $\Gamma_{\pi} \approx 10$
MeV is still small compared to the width of $\sim$ 150 MeV
expected for a two-body decay  of the \d into the $NN^*$ channel with 
subsequent pionic decay of the $N^*$ resonance. This channel is not included
in the present calculation because it is closed for the physically 
interesting \d masses around 2100 MeV. In the case that the 
\d mass is close to the experimental one,  
small pionic decay widths are obtained. Using the \d 
wave functions and parameters of sets 3 \& 5 ($b_6$'s are large), 
the calculated $\Gamma_{\pi}$'s are both 0.04 MeV which is 
an order of magnitude smaller than the experimental $\Gamma_{\pi}^{\rm exp}$.
We have also discussed that if one could obtain a \d mass of 
$\simeq 2100$ MeV with a size parameter not much different from that of a 
free-baryon, it would be possible to reproduce the empirical pionic decay 
width $\Gamma_{\pi}^{\rm exp}$.

	In this paper the role of the intermediate 
$s^4p^2[42]_X$-configuration in addition to the dominant $s^6[6]_X$-one 
is investigated and found to play an important role in the decay  width 
for some parameter sets. The inclusion of the $s^4p^2[42]_X$-configuration 
reduces the magnitude of the decay width compared with a calculation  
using only the $s^6[6]_X$-configuration. It is also shown that the nonstatic 
correction in the decay operator changes the width even 
at low-momentum transfers and that the form factor at the decay
vertex reduces the decay width for the parameter sets involving
high-momentum transfers. 

    We conclude that the experimentally observed small pionic decay 
width $\Gamma_{\pi}^{\rm exp} \simeq 0.5$ MeV of the \d-dibaryon can be 
explained by the microscopic shell model wave function and the 
single-quark transition operator provided that the model 
can reproduce the mass $M_{d^\prime}$ correctly and the size of \d is 
not very different from the size of a free baryon. Finally, we find 
it important and necessary to go beyond the plane wave approximation 
and to solve the $\pi NN$ final state wave function exactly 
within the present model in order to obtain an improved pionic decay width 
of the \d-dibaryon.

\vskip 15mm

\centerline{ Acknowledgement}

\vskip 3mm
	One of the authors (K.I.) acknowledges the warm hospitality 
extended to him during his stay at the Institute for Theoretical Physics, 
University of T\"ubingen where this work has been mostly completed.
He also wishes to thank the Ministry of Education, Science, Sports
and Culture (Japan) for finanicial support which made this stay at the 
University of T\"ubingen possible. 

\vfil\eject

Appendix

\vskip 3mm

The pionic decay width of the \d-dibaryon is expressed in our model as 

\bd
 \Gamma_\pi = 3 \ \Gamma_{\pi^-} \quad ,  
\ed
 
\beas
 \Gamma_{\pi^-} = & \displaystyle \int _0^{q_{\rm max}}& q^2 dq 
 \frac {f_{\pi q}^2} 
 {4\pi m_\pi^2} \frac {2M_N} {2M_N + E_\pi} \cdot 2k_0 \cdot ({\rm F.F.} ) 
  \left( \frac {2 (\frac {b} {b_N})} { 1+ (\frac {b} {b_N})^2 } \right)^{12}
 \\
 & \times & \left\vert \ \sqrt {\frac 2 9} \phi_{00}(q,b) \frac {10} {9} 
 \frac {1} {\sqrt {5}} [{\rm NSCI.}] \ bk_0^2 \exp[{- \frac 5 {24} b^2k_0^2 }] 
 \right. \\
 & & \left. - \sqrt {\frac 2 {25}} \phi_{10}(q,b) \frac 3 {10} 
 \frac 1 {\sqrt {10}} 
( 1 - \frac {2\sqrt {3}} {9} ) \ bk_0^2 \exp[{- \frac 5 {24} b^2k_0^2}] \ 
 \right\vert ^2 
\eeas
 
\noindent
where

\bd
  ({\rm F.F.}) = \frac {\Lambda^2} { \Lambda^2 + {\mbox{\boldmath $k$}}_0^2 } 
\quad ,
\ed

\bea
  [{\rm NSCI.}] &=& 1 + \frac {m_\pi} {2m_q}  \nonumber \\
   &- & (\frac {m_\pi} {2m_q}) \left \{ \frac 1 2  
    + \frac { \ell_R (\ell_R +1) - \ell_L (\ell_L +1) } {2} \right \}
    \frac {12} 5 \frac 1 {(bk_0)^2}   \nonumber \\
   & & ; \quad  (\ell_R = 1, \ \ell_L = 0 )  \quad \quad , \nonumber
\eea

\bd
  E_\pi = \sqrt {m_\pi^2 + k_0^2} \quad ,  
\ed

\bd 
  \phi_{00}(q,b) = \sqrt { \frac { 2^3 \sqrt {6} b^3} {\sqrt \pi 3^2}} 
  \exp[{- \frac {b^2 q^2} 3 }] \quad , 
\ed

\bd 
  \phi_{10}(q,b) = \sqrt { \frac { 2^2 \sqrt {6} b^3} {\sqrt \pi 3} }
  (1 - \frac 4 9 b^2q^2 ) \exp[{- \frac {b^2 q^2} 3 }] \quad , 
\ed
 
\bd 
  k_0 = \left[ 4M_N \left\{ (M_{d^\prime} - \frac {q^2} {M_N} ) 
  - \sqrt { (M_{d^\prime} - \frac {q^2} {M_N})^2 
   - (M_{d^\prime} - 2 M_N - \frac {q^2} {M_N})^2 + m_\pi^2 } \ \right\} 
  \right]^{1/2} \quad , 
\ed

\bd
  q_{\rm max} = \sqrt { M_N( M_{d^\prime} - 2M_N -m_\pi) } \quad . 
\ed

	In the above expression the h.o. size parameter $b \ (b_N)$ denotes 
that of the \d-dibaryon ( free nucleon ).  (F.F.) represents the 
vertex form factor and [NSCI.] the nonstatic correction effect associated 
with Eq. (10).

\end{document}